# Physics-informed self-supervised generation of digital brain MRI phantoms from weighted images using differentiable MRI simulation

Authors:

Kseniya Belousova[1], Anna Konanykhina[1], Zilya Murzina[1], Walid Al-Haidri [1], Ekaterina Brui[1]

Affiliations:

[1] School of Physics and Engineering, ITMO University, Saint-Petersburg, Russia

Corresponding author:

Walid Al-Haidri
School of Physics and Engineering, ITMO University
9 Lomonosova St., office 2426 St. Petersburg 191002 Russia
walhaidri@itmo.ru

# Abstract

**Purpose:** To develop a physics-informed, self-supervised framework for generating digital brain MRI phantoms directly from conventional weighted MR images without requiring ground-truth parametric maps or anatomical segmentation.

**Methods:** The framework predicts T1, T2, and proton density (PD) maps from T1-, T2-, and PD-weighted images and reconstructs the input images through an MRI signal model. Three generative architectures (variational autoencoder (VAE), generative adversarial network (GAN), and flow-based model) were compared. Models were pretrained on 3,739 synthetic brain slices generated using digital phantoms and an analytical MRI signal model, followed by fine-tuning on 90 real brain slices from three healthy volunteers. The best-performing architecture was subsequently fine-tuned using the differentiable MR-Zero numerical MRI simulator and evaluated on 30 held-out real slices.

**Results:** The flow-based model demonstrated the highest overall performance and preserved fine anatomical details better than the VAE and GAN. After analytical-model fine-tuning, it achieved MS-SSIM values of 0.955-0.985 and PSNR values of 26.68-32.63 dB across T1-, T2-, and PD-weighted images. Fine-tuning with MR-Zero increased T1-weighted reconstruction quality from 0.955 to 0.977 (MS-SSIM) and from 26.68 to 30.60 dB (PSNR), and provided high robustness of metrics across different MR image weightings. The resulting digital phantoms also enabled simulation of images using previously unseen acquisition protocols.

**Conclusion:** The proposed framework enables physics-informed generation of reusable digital brain MRI phantoms from weighted images using limited real-world data. Combining synthetic pretraining with differentiable numerical MRI simulation provides a practical approach for physically grounded MRI data augmentation without requiring reference parametric maps.

**Keywords:** MRI; digital phantom; self-supervised learning; MR-Zero; differentiable MRI simulation

# 1. Introduction

Phantoms play an essential role in quality control, calibration, research, and clinical training across various medical imaging modalities [1]. While early advancements relied on physical phantoms, the development of numerical simulation techniques has led to the emergence of digital phantoms: virtual, computationally generated models that replicate tissue properties [2]. In magnetic resonance imaging (MRI), a digital phantom comprises at least a baseline set of multi-parametric tissue maps: a spin–lattice relaxation time (T1) map, a spin–spin relaxation time (T2) map, and a proton density (PD) map. Additionally, it may incorporate other parameter layers representing alternative tissue properties [3].

Recently, the integration of AI in medicine has renewed the significance of digital MRI phantoms, which offer a powerful solution to the chronic scarcity of annotated clinical data - a critical bottleneck in training deep learning models for segmentation, reconstruction, and MRI quality enhancement [4]. While neural networks dominate MR image analysis, acquiring the required large datasets is severely constrained by high scanning costs and patient confidentiality. A single digital phantom bypasses these limitations, enabling the generation of diverse, realistic contrast images via different pulse sequences [5], which eliminates the need for repeated volunteer scanning or physical phantoms [6]. This capability renders digital phantoms indispensable for scalable, privacy-compliant synthetic data generation in AI-driven radiology [7, 8]. Consequently, the efficient creation of realistic digital MRI phantoms has emerged as a vital research challenge.

Traditionally, the constituent parameter maps for digital MRI phantoms were acquired via quantitative MRI (qMRI) techniques. These approaches require multiple scans of the same anatomical region with varied pulse sequence parameters, followed by non-linear voxel-wise fitting to derive the maps [9]. However, this approach is highly time-consuming and sensitive to patient motion during the extended acquisition. To address this, fast single-scan techniques have been developed for simultaneous multi-parametric mapping. The most notable is magnetic resonance fingerprinting (MRF) [10], which matches a pseudo-random temporal signal evolution from each voxel to a precomputed Bloch equation dictionary. Other single-scan alternatives include IR-TrueFISP [11], QRAPMASTER [12], and MPME [13]. Nevertheless, a major clinical bottleneck of all these fast mapping approaches is their strict reliance on dedicated, non-standard pulse sequences unavailable on routine scanners.

Several studies have integrated deep learning with MRF reconstruction by replacing dictionary matching with convolutional neural networks or fully connected architectures [15,16]. While these methods considerably accelerate parameter estimation, they still rely on supervised training with large simulated dictionaries, requiring extensive Bloch simulations and limiting flexibility when protocols are modified. Consequently, they alleviate computational costs during reconstruction but do not eliminate the dependence on specialized MRF pulse sequences. Alternatively, researchers have used Denoising Diffusion Probabilistic Models (DDPMs) to synthesize parametric T1 maps [17] and U-Net architectures to generate T2 maps from weighted images [18]. Nevertheless, these approaches cannot produce the comprehensive suite of relaxation maps required to construct a complete digital phantom; furthermore, they strictly depend on ground-truth reference maps for training, which are inherently difficult to acquire in clinical practice. To achieve simultaneous multi-parametric synthesis (T1, T2, and PD) from standard weighted images, a CNN-based model was proposed utilizing tissue maps derived from anatomical masks [17]; however, relying on such masks is impractical in routine clinical workflows. Another strategy employs pre-derived maps obtained via Multi-Dynamic Multi-Echo (MDME) techniques to generate weighted MR images through an analytical signal model, using a U-Net to integrate input images and relevant acquisition parameters to guide the synthesis process [3]. Nevertheless, like previous approaches, this framework fundamentally depends on access to either synthetic or real reference maps, which constitutes a significant clinical limitation.

A more promising approach involves self-supervised strategies that eliminate the need for ground-truth reference maps or anatomical masks. For instance, a self-supervised framework was proposed in [9] using a U-Net to predict multi-parametric maps (T1, T2, and PD) from three input modalities, optimizing the network via a physics-based reconstruction loop. Similarly, a latent diffusion model combined with a multimodal variational autoencoder (VAE) was introduced in [19], incorporating pulse-sequence parameters into the encoder. Despite being trained on real MRI data, this latter approach exhibits a notable limitation: the predicted parameter values do not always remain within physically plausible ranges. Furthermore, both frameworks [9, 19] rely on simplified analytical equations to model the MR signal for computational speed. Because these analytical models account for only a limited subset of pulse-sequence parameters, their physical fidelity is restricted.

In this study, we present a physics-informed, self-supervised framework for the simultaneous synthesis of multi-parametric tissue property maps (T1, T2, and PD) from standard weighted images, enabling the creation of high-fidelity digital phantoms without relying on ground-truth reference data. Rather than utilizing only simplified analytical models, our approach incorporates a numerical MRI simulator as an end-to-end consistency constraint to enforce physical plausibility. Within this paradigm, we implement and systematically compare three distinct generative architectures - Variational Autoencoders [19], Generative Adversarial Networks [20], and Flow-based models [21]. Furthermore, to mitigate clinical data scarcity and bridge the sim-to-real gap, we investigate a two-stage training pipeline that leverages large-scale synthetic phantom pre-training followed by fine-tuning on limited clinical scans. Finally, the framework is validated by constructing digital phantoms from the predicted maps and evaluating their capacity to synthesize realistic weighted MR images under varying imaging conditions.

# 2.Methods

## 2.1 Datasets

This study employed two datasets for model development and evaluation: (i) a synthetic MRI dataset and (ii) a real-world MRI dataset. Both datasets contained T1-, T2-, and PD-weighted images acquired or simulated using identical pulse-sequence parameters. Three two-dimensional single-slice turbo spin-echo (TSE) pulse-sequences were utilized. All protocols shared the same field of view (250 × 250 $mm^2$) and spatial resolution (1.95 × 1.95 × 5 $mm^3$). The protocols were: for T1-weighted (TR = 700 ms, effective TE = 7.65 ms, echo train length = 10), for T2-weighted (TR = 6.0 s, effective TE = 76.5 ms, echo train length = 10), and for PD-weighted (TR = 6.0 s, effective TE = 6.69 ms, echo train length = 10). The matrix size of all images was 128 × 128 pixels.

### 2.1.1 Synthetic MRI dataset

Synthetic MRI data were generated using digital brain phantoms, an MRI signal simulator based on the Bloch equations, and Pulseq-compatible [22] pulse sequences.

**Phantom Generation**

Digital phantoms were constructed using 12 anatomical models from the BrainWeb repository [23]. Each anatomical model contained more than 300 brain slices. Every slice was represented by a segmentation mask comprising 11 tissue classes (Supporting Information Table S1). The tissue masks were assigned with the corresponding T1 and T2 relaxation times and PD values [2, 24, 25]. To introduce inter-slice physiological variability, tissue parameters were sampled from normal distributions centered at the reported mean values with a standard deviation equal to 10% of the mean. Using this approach, 3,839 digital phantoms (brain slices) were generated, each associated with corresponding T1, T2, and proton-density maps.

**MRI Simulation**

A TSE pulse sequence diagram was designed in PyPulseq open-source software tool [22]. It started with a 90° excitation sinc-shaped pulse (2,816 ms duration, 1.36 kHz bandwidth [26]) applied with a slice-selective gradient, followed by a rewinding gradient to rephase the spins. Then a series of 180° refocusing sinc-shaped pulses (3,84 ms, 0.68 kHz [26]) were applied, each with its own slice- selective gradient. Between echoes, phase- encoding gradients were stepped to encode spatial information. Before the first readout, a pre- phasing gradient was used. To remove unwanted signals (FID and stimulated echoes), crusher gradients were placed along the slice and phase directions around each readout. For k- space filling, we used centric-out ordering: the echo at the effective echo time ($TE_{eff}$) was placed in the central area of k- space, while echoes acquired earlier or later were placed farther away. Using this architecture, we created three separate Pulseq sequence files (.seq)

– one for each weighting: T1-, T2-, and PD-w. The only differences were the key timing parameters (Table 1). Synthetic MR images were generated using KomaMRI [27], an open-source MRI simulation framework. The simulator accepts digital phantoms and .seq sequence files as inputs and generates raw MR signals through numerical solutions of the Bloch equations. Finally, these signals were sorted into k-space, and the MR images were reconstructed using a 2D inverse fast Fourier transform. The resulting synthetic MRI dataset contained 3,739 sets of T1-w, T2-w, and PD-w images derived from 11 anatomical models for training, and 100 sets of weighted images obtained from a separate anatomical model for testing.

### 2.1.2 **The real-world MRI data**

The real-world dataset was acquired from four healthy volunteers within our research team (mean age: 35 ± 5 years). MRI examinations were performed on a Siemens Magnetom Espree 1.5T scanner at Almazov National Medical Research Centre, with the approval of the Institutional Review Board. All participants provided written informed consent prior to imaging. Each volunteer contributed 30 brain slices. Data from three volunteers (90 slices) were assigned to the training set, while data from one volunteer (30 slices) were reserved for testing.

## 2.2 Data preprocessing

To reduce the domain shift between synthetic and real images, synthetic samples were corrupted with Gaussian noise parameterized to match the statistical properties of the real data. Standard data augmentation was applied to both sets, including elastic transformations and rotations at 10°, 15°, and 90°. For the GAN model, the images were kept at their original resolution of 128×128 pixels, whereas for the VAE and HierarchyFlow models, they were resized to 256×256 pixels, which is appropriate for these architectures. Preprocessing involved applying Min-Max intensity normalization. Specifically, input MR images were scaled to [−1,1] for the VAE and to [0,1] for the GAN and HierarchyFlow models.

## 2.3 Pipeline of Self-Supervised MRI Parametric Map Estimation

### 2.3.1 Overview of the Training Framework

The overall workflow is illustrated in Figure 1. In this framework, the original weighted MR images are used as inputs into a neural network that predicts the corresponding parametric maps. The predicted maps are then passed to an MR signal model, which incorporates pulse sequence parameters (analytical model) or a full pulse sequence diagram (numerical model) to synthesize new weighted images. The neural network is trained by computing a reconstruction loss between the original input images and the synthesized images produced by the MR signal model.

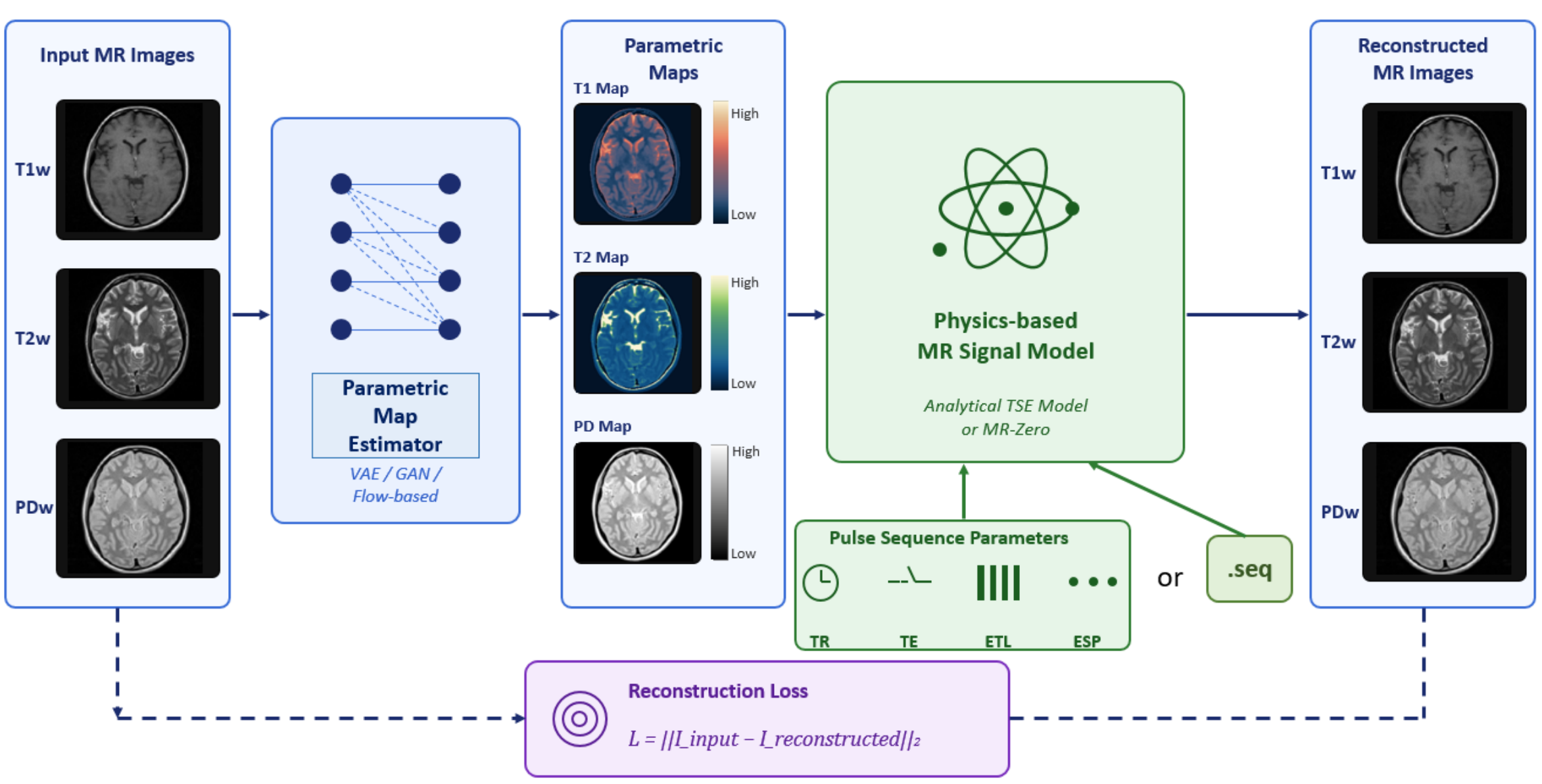

Figure 1. Overview of the proposed training pipeline. The pipeline takes weighted images (T1-, T2-, PD-w) as input to predict parametric maps via a neural network (VAE, GAN, or Flow-based). An MR simulator (Analytical or Numerical) then reconstructs weighted images from these maps. The model is optimized by minimizing the reconstruction loss between the input and the generated weighted images.

## 2.3.2 MR Signal Modeling

**Analytical Signal Model**

Since a full MRI simulator is computationally expensive, at the first step we utilized an analytical expression for the TSE signal model [28] (1). This model takes as input the pixelwise parameters (T1, T2, PD) together with the pulse sequence parameters specific to each contrast: repetition time (TR), echo time (TE), echo train length (ETL), and echo spacing (ESP). The model output is the simulated signal intensity at each pixel. This analytical formulation offers two major advantages: high computational efficiency and differentiability, both of which are essential for integrating the signal model into the neural network training process.

$$S = PD \times e^{\frac{-TE}{T2}} \times \left(1 - e^{\frac{-(TR - ETL \times ESP)}{T1}}\right) \quad (1)$$

**Numerical model**

MR-Zero simulation framework utilizes the differentiable phase distribution graph formalism, an extension of the extended phase graph concept [29]. This tool can be integrated directly into the training loop as a physics-based layer. This allows for end-to-end optimization, where gradients are propagated from the loss function through the MR simulator back to the neural network. The simulator required a digital phantom as input, along with .seq files, and generated raw k-space data as output, which were subsequently reconstructed into MR images

### 2.3.3 Models Training Strategy

This study investigated three neural network architectures - VAE, GAN, and flow-based models - for predicting digital phantom layers. The study pipeline was structured into three distinct phases.

1) First, each architecture was pretrained on a synthetic dataset using an analytical signal model and evaluated on both synthetic and real-world test sets.
2) Second, these pretrained networks were fine-tuned on a real-world training dataset, also utilizing the analytical signal model. This intermediate phase served exclusively to select the top-performing architecture based on the evaluation metrics (detailed in Section 2.5) obtained on the real-world test set.
3) In the final phase, only the selected top-performing architecture (initially pretrained on synthetic data) was fine-tuned on the real-world training dataset using the MR-Zero simulator as the numerical signal model, followed by evaluation on the real-world test set. Consequently, the final model combined pretraining on synthetic data via the analytical model with fine-tuning on real-world data via the numerical simulator - a strategy explicitly designed to mitigate the computational costs of the training procedure.

The training hyperparameters for each mode are summarized in Supporting Information Table S2. Adam was used as an optimizer in all cases.

## 2.4. Studied architectures:

### 2.4.1 Variational Autoencoder

We employed the multimodal VAE described in [19] as a baseline approach. This method handles multi-modal MR images, accounts for pulse sequence parameters, and supports self-supervised training guided by an analytical signal model. The multimodal VAE includes a shared convolutional encoder, which encodes each input modality independently into unimodal latent distributions. Each input image is associated with its specific pulse sequence parameters, which are used in adaptive group normalization within the residual blocks of the encoder. To fuse unimodal latent distributions into a joint multimodal distribution, we employ a product-of-experts approach. Assuming Gaussian unimodal distributions with encoder-predicted means and standard deviations, their product remains Gaussian. For decoding, a convolutional network is used: its outputs are parametric maps which pass through an exponential function (to avoid negative values) and then feed into the MR signal model to generate a weighted image. The loss function comprises a combination of $L_2$ reconstruction loss with perceptual loss, a patch-wise adversarial loss, and a KL-regularization loss on the latent distribution.

**Adaptation for digital phantoms**

For computing the reconstruction loss, we used unscaled images as targets, following the original work. However, in contrast to the original work, outputs from the analytical TSE model lie in the range [0, 1]. Therefore, reconstructed images were rescaled to the scale of the target images. The training scheme of this model is illustrated in the figure 2.

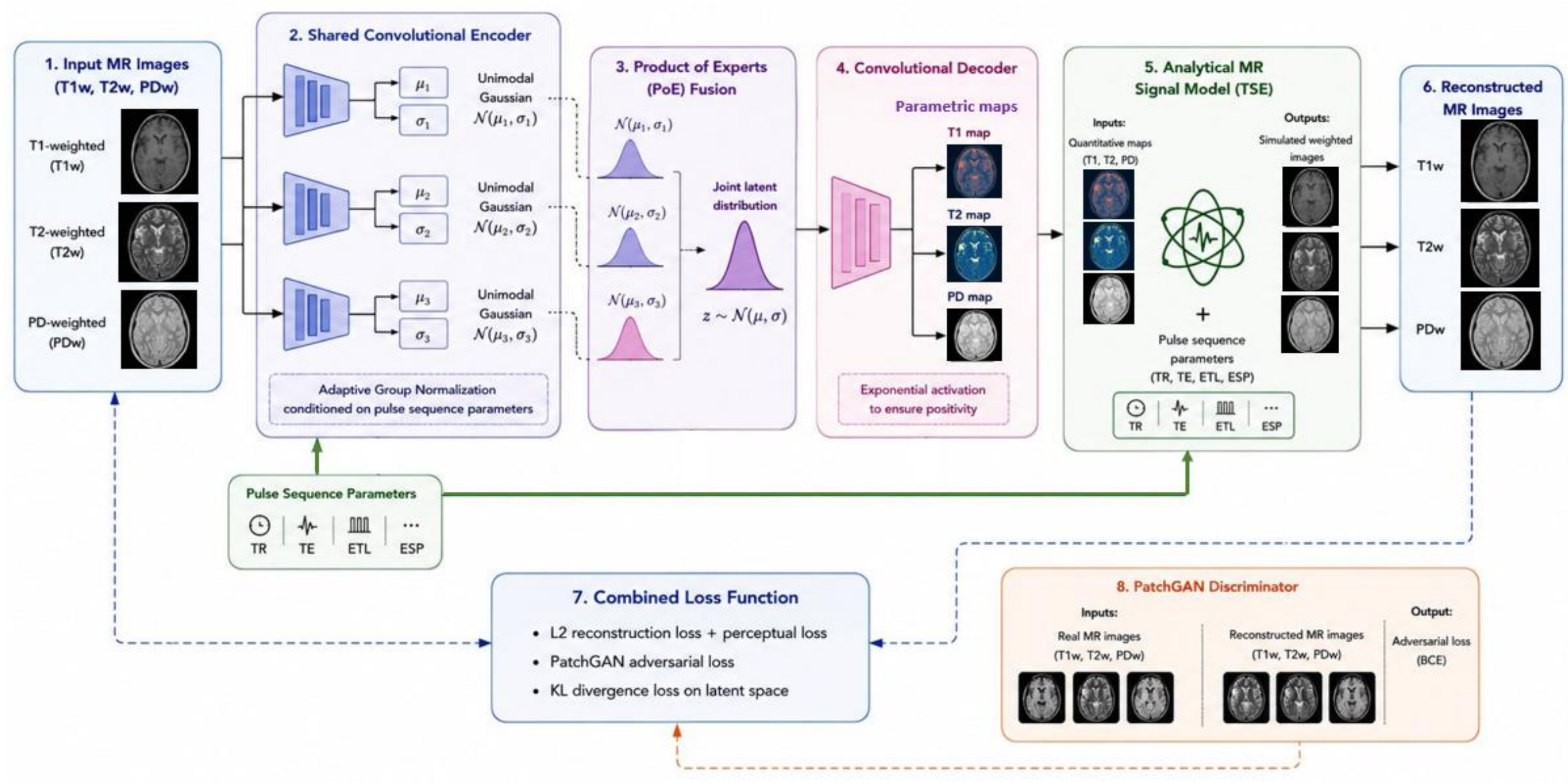

Figure 2. Overview of the multimodal variational autoencoder framework. T1-, T2-, and PD-weighted MR images are independently encoded into unimodal latent distributions using a shared convolutional encoder conditioned on pulse sequence parameters. The distributions are fused using the product-of-experts approach, and a convolutional decoder generates parametric T1, T2, and PD maps. These maps are subsequently used with an analytical MR signal model to reconstruct the corresponding weighted MR images. Training combines reconstruction and perceptual losses, PatchGAN adversarial loss, and KL-divergence regularization.

### 2.4.2 Flow-Based model

A flow-based model Hierarchy Flow for unpaired, high-fidelity image-to-image translation [21] was used as the second model. The original framework maximizes content preservation and structural fidelity during cross-domain synthesis, maintaining fine anatomy while avoiding conventional flow-model checkerboard artifacts. It incorporates a Hierarchical Coupling Layer base block, enabling multi-scale reversible transformations without spatial squeezing. Training requires source images (three-channel weighted image arrays) and target images (three-channel parametric maps) to transfer target style to the source. Importantly, the target images need not be content-aligned with the sources. In our setup, we use a single synthetic parametric map set (one T1 map, one T2 map, one PD map) as the target for training on both synthetic and real datasets.

**Adaptation for digital phantoms**

During the forward pass (Figure 3), hierarchical coupling layers successively encode the source images into a latent source feature representation. The target image (parametric map) is processed by a Style Network to extract style features, which are then injected into the source features via Adaptive Instance Normalization (AdaIN) [30]. The network is then inverted (run in reverse) to yield translated images. In the original methodology, a composite loss combining content and aligned style terms was employed. In our adaptation, the network's outputs are parametric maps with a range of values from 0 to 1. Next, the maps are scaled to characteristic (T1, T2, PD) ranges and fed into the MR signal model to generate new weighted images. Similarly to the VAE pipeline, an MSE loss between original and generated images is computed, accounting for their dynamic ranges.

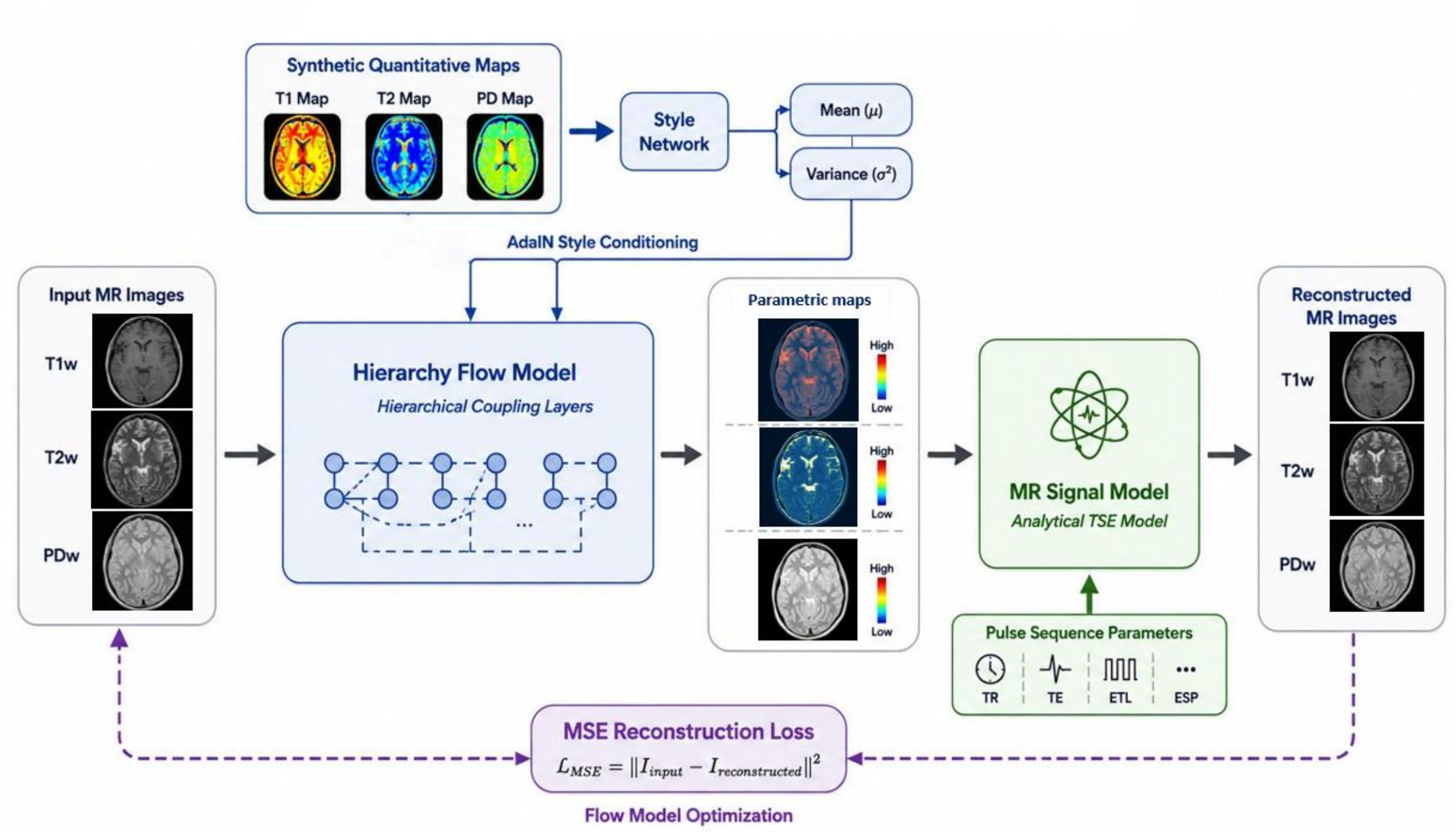

Figure 3. Flow-based model training scheme. The model consists of two Hierarchical Coupling Layers. It takes a three-channel array of weighted images (T1-, T2-, PD-w) as input, along with the mean and variance parameters derived from a single sample of synthetic parametric

maps using Style Net. The model synthesizes parametric maps (T1, T2, PD), which are then fed into an MR signal model to generate new weighted images for Mean Square Error (MSE) loss calculation.

### 2.4.3 Generative Adversarial Network (GAN)

As the third alternative, we selected a GAN model as described in [20]. This approach was chosen due to its demonstrated ability to generate high-quality images characterized by realistic structural features and texture fidelity [20].

The model comprises two components: generator and discriminator. The generator follows an encoder–decoder architecture with skip-connection blocks and closely resembles a U-Net structure. The discriminator is implemented as a PatchGAN, which divides the input image into patches and classifies each patch as either real or synthetic; the outputs are then averaged across all patches. This design specifically targets high-frequency structural components and may be regarded as implementing a form of texture/style loss, whereas the low-frequency corrections are governed by an L1 loss function.

**Adaptation of GAN for digital phantoms**

In contrast to the original implementation, the output activation function of the generator was replaced with the softplus activation function which is a smooth approximation to the ReLU function. As a result, the generated parametric maps are constrained to be strictly non-negative and are subsequently rescaled to the appropriate physical ranges based on the maximum expected values of T1, T2, and PD (Figure 4).

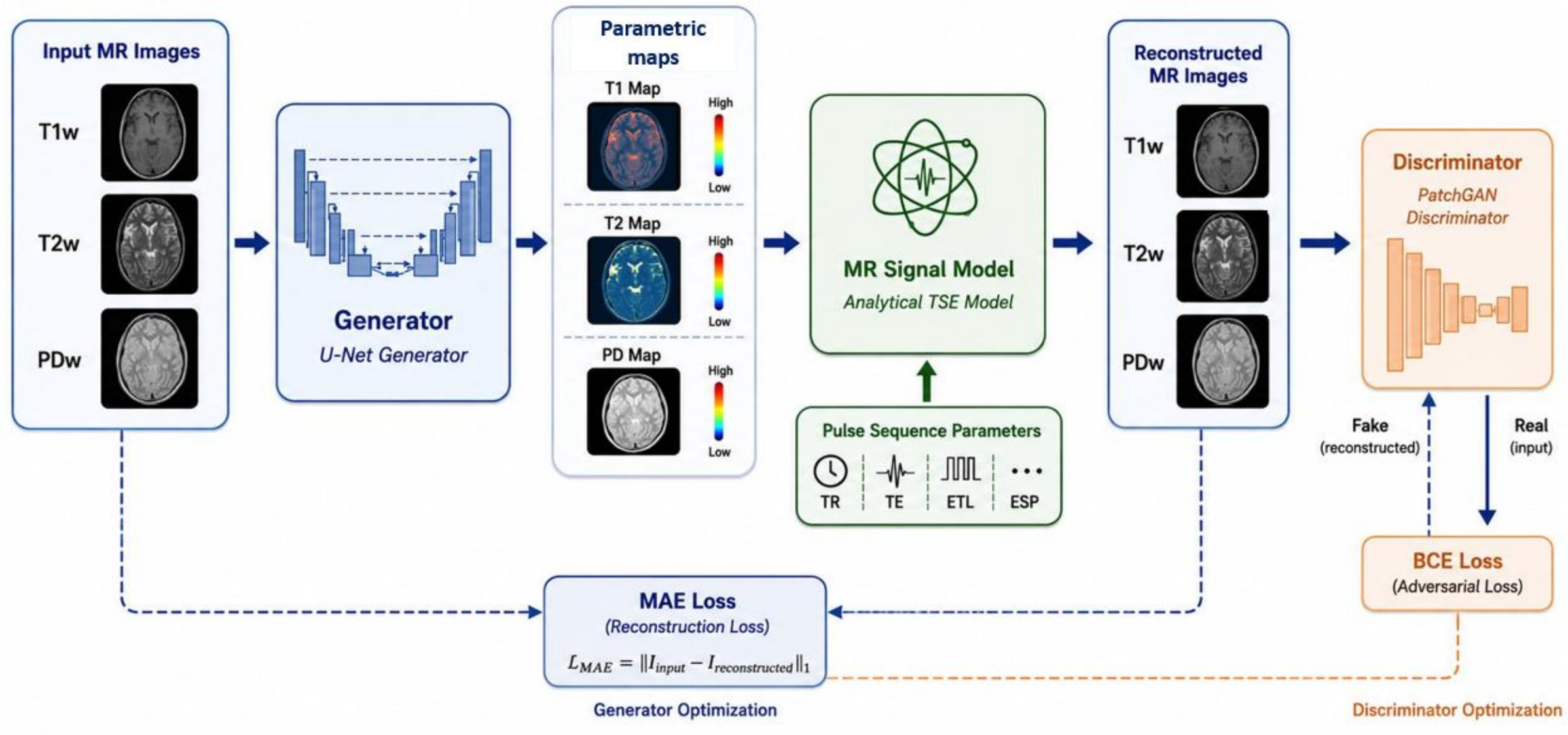

Figure 4. GAN training scheme. The generator takes weighted images as input and generates parametric maps, which are then fed into an MR signal model to produce new weighted images. These generated weighted images are passed to the discriminator. The discriminator is trained using Binary Cross-Entropy (BCE) loss, while the generator is trained using Mean Absolute Error (MAE) calculated between the original and generated weighted images.

The total loss function comprises the discriminator's binary cross-entropy loss and the Mean Absolute Error ($L_1$ loss) between the original weighted images and the synthesized weighted images. In computing the $L_1$ loss, the value ranges of both the original images and the generated images (from the generated maps) were accounted for, in accord with the aforementioned approach.

## 2.5 Quality evaluation

Model performance was evaluated by comparing the original MR images and the reconstructed images generated by the MR signal model from the predicted parametric maps. For quantitative assessment, we employed the Multi-Scale Structural Similarity (MS-SSIM) [31] (2) and Peak Signal-to-Noise Ratio (PSNR) [32] (3) metrics:

$$MS-SSIM(x,y) = \left(\prod_{j=1}^{M-1}\left[c_j(x,y)s_j(x,y)\right]^{w_j}\right) \times [l_M(x,y)]^{w_M} \quad (2)$$

$$PSNR = 20 \cdot \log_{10}\left(\frac{MAX_I}{\sqrt{MSE}}\right) \quad (3)$$

$$MSE = \frac{1}{MN}\sum_{n=0}^{M}\sum_{m=1}^{N}[x(n,m) - \hat{x}(n,m)]^2,$$

where x and y represent the reference and reconstructed images, respectively; $\hat{x}(n,m)$ denotes the pixel value of the reconstructed image at coordinates $(n,m)$; and $MAX_I$ is the maximum possible pixel intensity of the image.

# 3. Results

## 3.1 Performance of the models using an analytical MR signal model for training

As the first stage of model training employed synthetic data, for which we had created the ground truth (GT) parametric maps, it was possible to calculate MAE metric between GT and generated maps (Table 1).  The flow-based model achieved the lowest MAE for the T1 and T2 maps, while the GAN demonstrated the lowest MAE for the PD maps. For all three models, the highest MAE values are observed for T1 maps. Visual inspection of the generated maps (Figure 5) revealed distinct characteristics of the three models. The parametric maps generated by the VAE exhibited noticeable blurring and a loss of fine anatomical details. The GAN-generated maps demonstrated a high level of detail; however, they also contained structures that were absent from the original maps. The flow-based model produced maps with high anatomical fidelity and preserved fine details. However, the generated T1 values were lower than those in the GT maps, while the maximum T2 values were higher than the corresponding GT values.

Table 1. The mean absolute error (MAE) between the original synthetic parametric maps and those generated using different models.

| | VAE (mean±std) | GAN (mean±std) | Flow-based (mean±std) |
|---|---|---|---|
| T1 map | 0.204 ± 0.008 | 0.178 ± 0.011 | 0.149 ± 0.010 |
| T2 map | 0.094 ± 0.003 | 0.040 ± 0.003 | 0.040 ± 0.003 |
| PD map | 0.089 ± 0.002 | 0.061 ± 0.002 | 0.072 ± 0.001 |

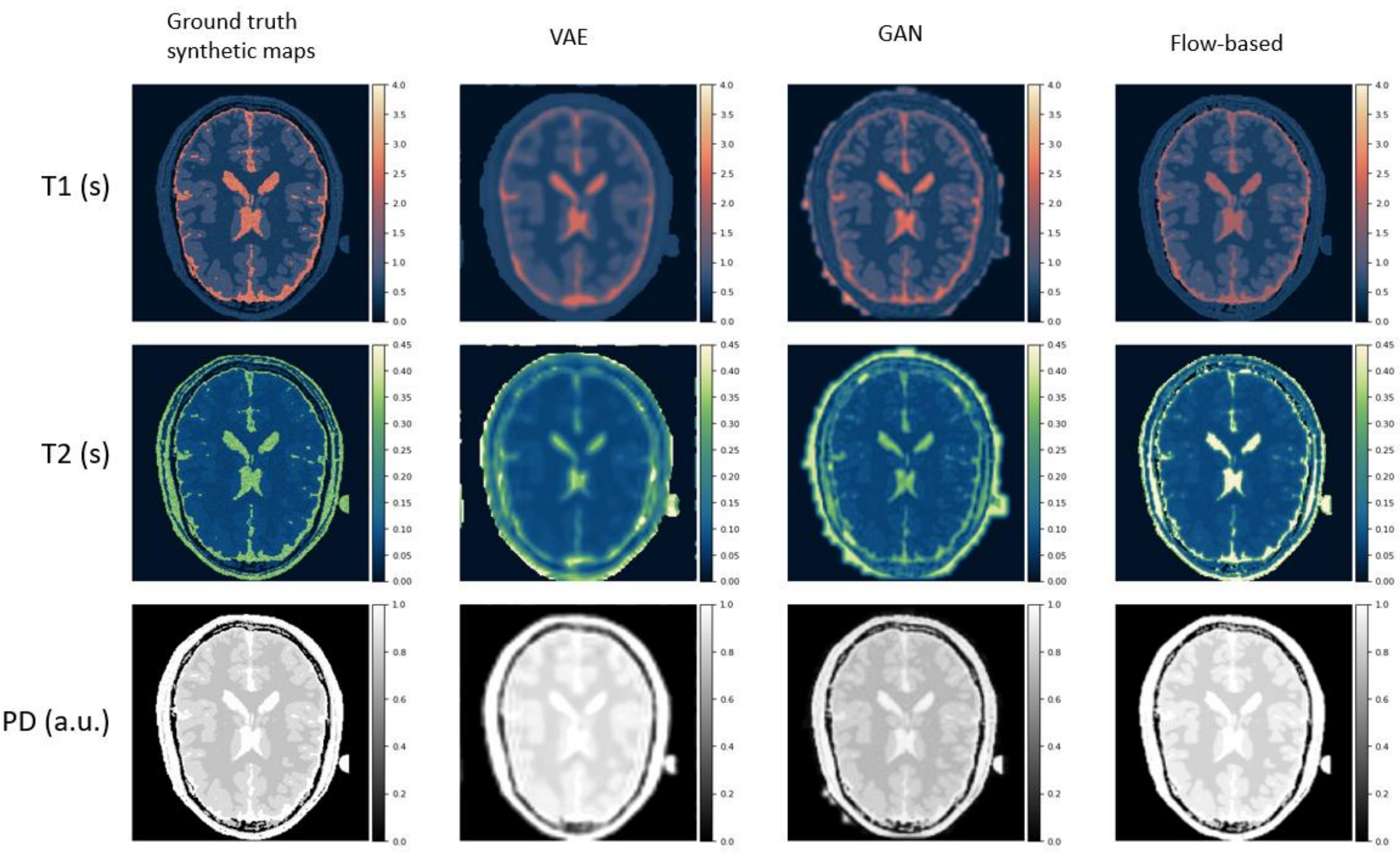

Figure 5. Examples of parametric maps: synthetic ground truth maps and those generated using VAE, GAN, Flow-based models.

For the evaluation on real-world data, direct comparison of the generated parametric maps with ground-truth maps was not available. Therefore, indirect metrics were used to assess the fidelity between the input images and the images reconstructed from the generated parametric maps. The quality metrics presented in Table 2 quantify structural similarity and pixel-wise accuracy between the reference input images and the images reconstructed using the analytical MR signal model. The results are presented for the models before and after fine-tuning on real-world data.

Table 2. Performance of VAE, GAN, and Flow-based models on the real-world test set, trained using an analytical MR signal model. MS-SSIM and PSNR metrics were calculated between input images and those reconstructed via the analytical MRI signal model using the generated maps. The results are presented before fine-tuning on a real dataset and after it.

| | | VAE | | GAN | | Flow-based | |
|---|---|---|---|---|---|---|---|
| Image type | Metric | Pretrained on synthetic data | Fine-tuned on real-world data | Pretrained on synthetic data | Fine-tuned on real-world data | Pretrained on synthetic data | Fine-tuned on real-world data |
| T1-weighted | MS-SSIM (mean±std) | 0.521 ± 0.063 | 0.917 ± 0.018 | 0.960 ± 0.005 | 0.962 ± 0.013 | 0.954 ± 0.008 | 0.955 ± 0.005 |
| | PSNR (mean±std) | 14.46 ± 0.60 | 29.71 ± 1.27 | 27.07 ± 1.15 | 25.90 ± 1.90 | 25.51 ± 1.34 | 26.68 ± 1.17 |
| T2-weighted | MS-SSIM (mean±std) | 0.827 ± 0.034 | 0.964 ± 0.017 | 0.899 ± 0.038 | 0.983 ± 0.006 | 0.973 ± 0.021 | 0.985 ± 0.007 |
| | PSNR (mean±std) | 19.38 ± 0.80 | 31.37 ± 1.90 | 20.29 ± 1.18 | 29.06 ± 1.90 | 30.24 ± 1.75 | 32.63 ± 1.28 |
| PD-weighted | MS-SSIM (mean±std) | 0.681 ± 0.073 | 0.961 ± 0.010 | 0.771 ± 0.050 | 0.965 ± 0.0129 | 0.979 ± 0.004 | 0.975 ± 0.006 |
| | PSNR (mean±std) | 17.88 ± 0.99 | 34.17 ± 2.15 | 19.24 ± 0.67 | 28.77 ± 1.54 | 32.89 ± 0.95 | 32.52 ± 0.99 |

As shown in Table 2, the flow-based model achieved the best performance on the real-world test set before fine-tuning on real-world data. It also demonstrated the highest performance after fine-tuning, although the improvement in performance was relatively small. In contrast, the VAE and GAN showed lower performance before fine-tuning and a more pronounced improvement in the quality metrics following fine-tuning on the real-world dataset.

The images reconstructed from the parametric maps generated by the VAE exhibited pronounced blurring, a loss of fine anatomical details, and minor changes in contrast (Figure 6 A). The GAN preserved fine anatomical details well; however, changes in contrast, particularly in the PD-weighted images, were noticeable. In addition, isolated regions with abnormally high pixel values were observed in the T1-weighted images. The images reconstructed from the flow-based model output showed relatively minor contrast changes, which were more pronounced in the CSF region of the T1- and PD-weighted images. At the same time, fine anatomical details were well preserved.

Visual assessment of the generated parametric maps also revealed model-specific differences (Figure 6 B). The maps generated by the VAE exhibited pronounced smoothing of anatomical structures. In addition, the skull regions in the T1 and T2 maps contained non-zero values instead of the expected background values. The parametric maps generated by the GAN demonstrated a high level of anatomical detail and preserved fine structures; however, regions with underestimated values were observed in the CSF region in the T1 maps. At the same time, the parametric maps generated using flow-based model demonstrate high anatomical detail, the absence of blurring, and underestimated regions.

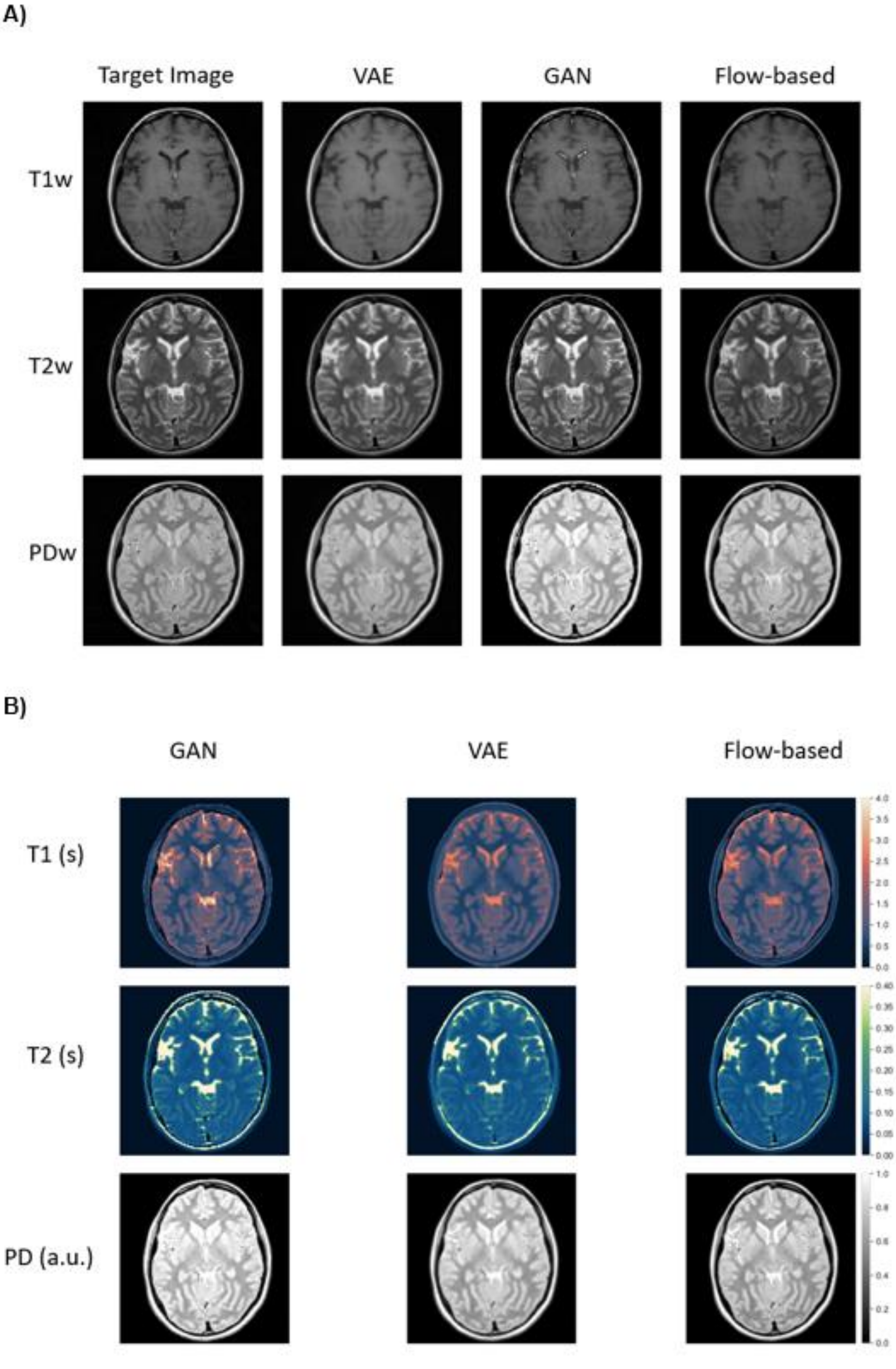

Figure 6. Model results: A – images reconstructed using an analytical MR signal and B – generated parametric maps. T1(s) - T1 relaxation time in seconds, T2(s) - T2 relaxation time in seconds, PD(a.u.) - proton density in arbitrary units, T1w - T1-weighted image, T2w - T2-weighted image, PDw - PD-weighted image.

Overall, considering both the metrics and the visual characteristics of the reconstructed weighted images and parametric maps, the flow-based model demonstrated the most satisfactory performance.

## 3.2 Performance of the flow-based model fine-tuned on MR-Zero MRI simulator

The flow-based model, that was initially pretrained for 10 epochs on synthetic data using the analytical MR signal model, was subsequently fine-tuned for an additional 7 epochs on a real-world dataset using the MR-Zero simulator. The resulting metrics are presented in Table 3. Examples of the generated parametric maps and reconstructed images obtained using the MR-Zero simulator are shown in Figure 7. For comparison, the results obtained after fine-tuning on the real-world dataset using the analytical MR signal model are also included in Table 3 and Figure 7 for the same slice.

The model fine-tuned on real-world data using the MR-Zero simulator achieved higher MS-SSIM and PSNR values for the T1-weighted images compared with fine-tuning using the analytical signal model. For the other two image weightings, MS-SSIM values were comparable between the two fine-tuning approaches, whereas PSNR values were higher after fine-tuning using the analytical MR signal model. Notably, fine-tuning with MR-Zero resulted in a more balanced distribution of the quality metrics across the three image weightings.

Visual assessment of the reconstructed images (Figure 7) demonstrated that the images obtained after fine-tuning with MR-Zero most closely matched the reference images in terms of contrast across all three weightings while maintaining high sharpness and preservation of anatomical details.

Table 3. Results for the flow-based model fine-tuning. Pre-training was conducted on synthetic data using an analytical MR signal model, followed by fine-tuning on real-world data using either analytical model or the MR-Zero simulator. Testing was performed on a real-world dataset.

| | T1w analytical signal model | T1w MR-Zero simulator | T2w analytical signal model | T2w MR-Zero simulator | PDw analytical signal model | PDw MR-Zero simulator |
|---|---|---|---|---|---|---|
| MS-SSIM (mean±std) | 0.955 ± 0.005 | 0.977 ± 0.004 | 0.985 ± 0.007 | 0.985 ± 0.002 | 0.975 ± 0.006 | 0.974 ± 0.0111 |
| PSNR (mean±std) | 26.68 ± 1.17 | 30.60 ± 1.21 | 32.63 ± 1.28 | 30.53 ± 1.38 | 32.52 ± 0.99 | 30.14 ± 1.21 |

To further assess the applicability of the synthesized digital phantoms beyond the acquisition protocol used for training, we used the MR-Zero simulator to generate images with pulse sequences not encountered during model optimization. Specifically, T1-weighted 2D FLASH, 2D turbo-FLASH, and 2D FLAIR .seq files were constructed in PyPulseq and simulated in MR-zero using the digital phantom generated by the flow-based model. The acquisition parameters were selected according to the recommended ranges for standardized clinical brain MRI protocols [33]. For all simulations, a 256 × 256 matrix was used. Despite the substantial differences in sequence timing and contrast mechanisms, the generated digital phantoms could be used to produce anatomically plausible MR images across all three acquisition protocols. Figure 7 (C) compares one of the original TSE images (T1-w) with the simulated images, illustrating the ability of the proposed approach to generate diverse contrasts from a single quantitative representation.

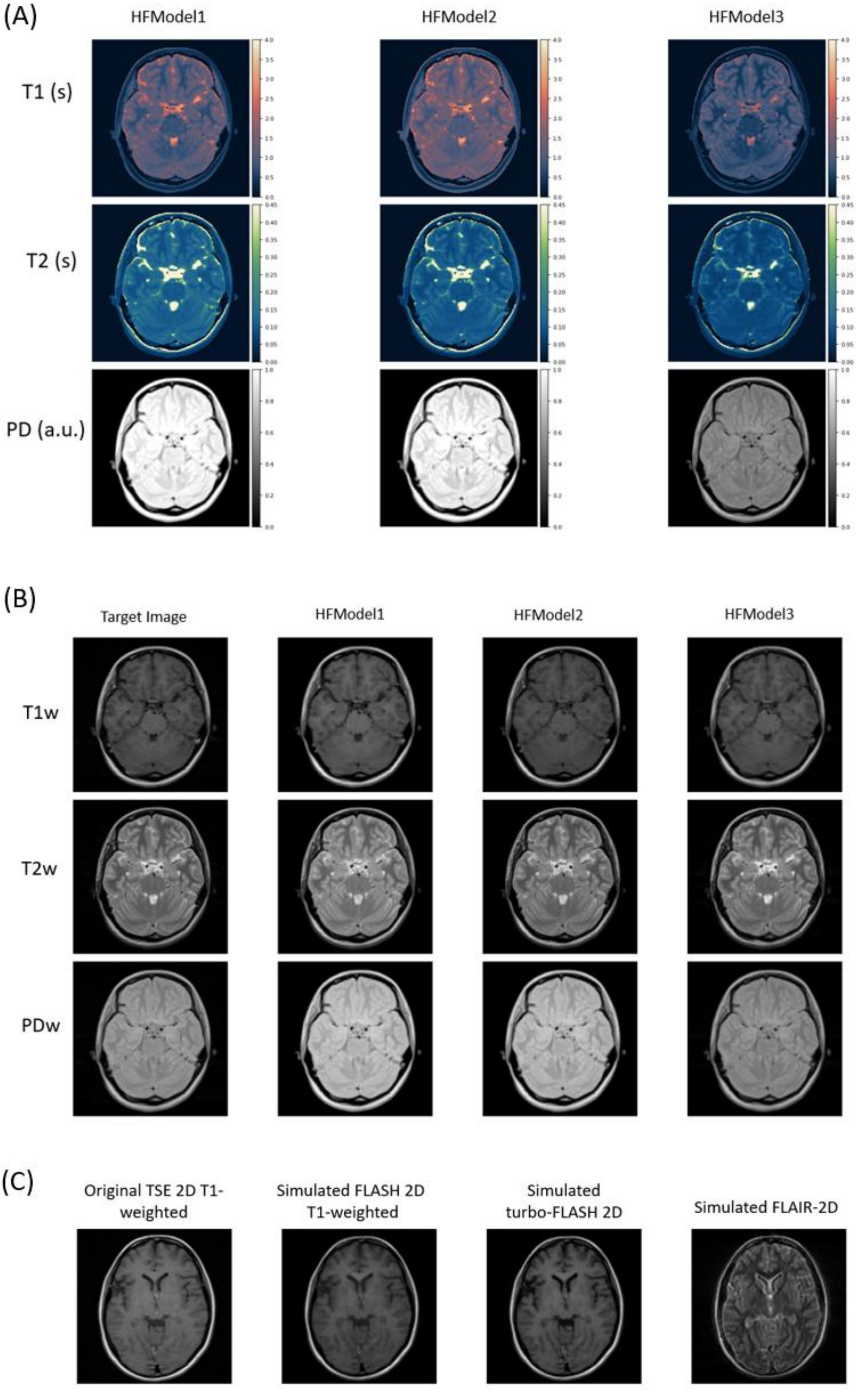
(A)
HFModel1
HFModel2
HFModel3
T1 (s)
T2 (s)
PD (a.u.)
(B)
Target Image
HFModel1
HFModel2
HFModel3
T1w
T2w
PDw
(C)
Original TSE 2D T1-weighted
Simulated FLASH 2D T1-weighted
Simulated turbo-FLASH 2D
Simulated FLAIR-2D

Figure 7. (A) Examples of parametric maps generated by the flow-based model under three training settings. HFModel1 – the model pre-trained on synthetic data using an analytic MR signal model, HFModel2 – the model pre-trained on synthetic and finetuned on real data using an analytic MR signal model, HFModel3 – the model pre-trained on synthetic and finetuned on real data using MR-Zero simulator. (B) Comparison between original TSE images and images generated using the flow-based model and the MR-Zero simulator. The flow-based model was trained using the MR-Zero simulator. HFModel1 – the model pre-trained on synthetic data using an analytic MR signal model, HFModel2 – the model pre-trained on synthetic and finetuned on real data using an analytic MR signal model, HFModel3 – the model pre-trained on synthetic and finetuned on real data using MR-Zero simulator. (C) Simulation of different brain MRI contrasts from a single digital phantom. From left to right: original T1-weighted TSE image and MR-Zero simulations of T1-weighted 2D FLASH (TR/TE/FA = 250ms/5ms/70°), turbo-FLASH-2D (TR/TE/TI/FA = 2100ms/4ms/1100ms/12°), and FLAIR-2D (TR/TE/TI/FA/ETL = 9000ms/110ms/2200ms/90°/180°/14). All simulated images were generated using a 256 × 256 matrix.

# 4. Discussion

In this study, we developed a two-stage, self-supervised neural network framework for generating realistic and physically accurate digital phantoms of the human brain. The proposed method successfully combines neural network pretraining on synthetic data via an analytical model with fine-tuning on real-world data via a numerical simulator - a strategy explicitly designed to mitigate the computational costs of the training pipeline. Our findings demonstrate that when utilized as a physical signal model during self-supervised fine-tuning, the MR-Zero numerical simulator outperforms the analytical signal model by providing higher robustness of metrics across different MR image weightings. Furthermore, among the three distinct architectures evaluated (VAE, GAN, and Flow-based models), the Flow-based architecture achieved superior results. Despite the limited volume of real-world training data, the models achieved high-quality performance metrics on the test set – an outcome largely enabled by synthetic data pretraining strategy. Crucially, the self-supervised approach eliminates the requirement for ground-truth quantitative maps. This shift fully distinguishes the proposed framework from previous studies that heavily rely on reference target data [17, 18], making our pipeline highly applicable to clinical scenarios where reference maps are fundamentally unavailable.

Our work is closely related to the physics-informed latent diffusion framework proposed by Lüpke et al. [19], from which we adapted the multimodal VAE architecture. The authors reported an averaged MS-SSIM of 0.9577 and PSNR of 30.89 dB, while the corresponding values in our experiments varied across contrasts, reaching MS-SSIM values of 0.917-0.961 and PSNR values of 29.71-34.17 dB for the fine-tuned VAE. Although these values are broadly comparable, direct quantitative comparison is limited by differences in datasets, pulse sequences, spatial resolution, and preprocessing. Importantly, our model was pretrained on 3,739 synthetic brain slices and subsequently fine-tuned using only 90 real-world training slices, whereas Lüpke et al. trained their VAE on real OASIS-3 MRI data. This highlights the potential of synthetic-data pretraining for reducing the amount of real MRI data required for physics-informed model training. Moya-Sáez et al. [34] demonstrated the feasibility of generating T1, T2, and proton-density maps from two routinely acquired weighted images (T1w and T2w) using a convolutional neural network trained on synthetic BrainWeb data. In contrast to our framework, the network was trained in a supervised manner using synthetic parametric maps as explicit reference targets. Lune et al. [35] recently proposed a self-supervised physics-guided framework for estimating T1, T2, and PD maps from conventional T1w, T2w, and FLAIR MRI. The model predicts parametric maps that are subsequently used to reconstruct the input MRI through an analytical signal model. However, the objectives and experimental settings of the study differ substantially: while van Lune et al. demonstrate the feasibility and robustness of self-supervised parametric (quantitative) mapping at clinical scale, our work investigates how the similar physics-guided principle can be combined with synthetic-data pretraining and numerical MRI simulation to construct reusable digital MRI phantoms from limited real-world data.

While analytical models may not capture complex pulse-sequence effects and are difficult to derive for diverse sequence types, numerical simulators can reproduce MRI signal formation based on magnetization dynamics. In this work, KomaMRI was used separately to efficiently generate the large-scale synthetic pretraining dataset, benefiting from its fast forward simulation and efficient memory management [27]. For fine-tuning, we transitioned to MR-Zero because its end-to-end differentiability allows integration into the backpropagation loop [29]. While previous studies have primarily employed MR-Zero for automated pulse sequence optimization [29, 36, 37], our work demonstrates its use as a physical signal constraint for self-supervised parametric map synthesis without target reference.

When evaluating the results, it can be observed that images reconstructed from the digital phantoms generated by the VAE and GAN models exhibited distinct artifacts. The VAE generated visibly smoother maps, consistent with previous reports of blurred outputs in VAE-based medical image synthesis [38]. GAN-based generation, while capable of producing visually detailed images, is known to be susceptible to localized intensity artifacts and hallucinated structures that are not supported by the source image [39, 40]. In contrast, the Flow-based model produced maps that yielded high image sharpness and anatomical precision. This superior performance is attributed to the inherent architecture of normalizing flows, which utilizes a series of reversible transformations and avoids spatial downsampling stages - a key advantage for preserving fine brain structures.

Despite these advantages, flow-based models can be computationally demanding due to their invertible architectures generally provide less dimensionality reduction than latent-variable models such as VAEs [41]. In our study, this limitation was partly mitigated by the compact architecture used, which contained approximately 1.0 million trainable parameters. The relatively small model size was particularly advantageous for training with the differentiable MR-Zero simulator, whose memory requirements substantially increase during forward simulation and backpropagation. Thus, the compact Flow-based architecture allowed us to retain detailed parametric maps while keeping the combined neural-network and physics-based training computationally feasible.

This study has several limitations. First, although our framework was validated using a specific set of three input pulse sequences (T1-w, T2-w, and PD-w TSE). The integration of an end-to-end differentiable Bloch solver ensures that the pipeline can be fine-tuned to any acquisition protocol, provided the input data physically captures T1-, T2- and PD-weighting and can be numerically simulated. Second, the generated synthetic images represent an idealized scenario lacking experimental B0 and/or B1+ field inhomogeneities and noise. It should be noted that at this stage we intentionally do not use the term “quantitative maps” for our digital phantoms, considering that the proposed pipeline does not account for the fields distributions, is not validated by ground-truth measurements, and cannot be used for precise clinical qMRI. However, as a future step, these fields variations can be incorporated as additional spatial layers within the digital phantom during simulation, while the realistic Rician noise can be efficiently applied during post-processing. Third, the current model is spatially constrained to a fixed resolution and trained exclusively on healthy brain anatomy. Future work will focus on scaling the framework to arbitrary voxel geometries, incorporating multi-component tissue properties like diffusion, and introducing pathological anomalies into the pre-training dataset to enhance clinical utility.

# 5.Conclusion

This work presents a method for generating digital phantoms directly from weighted MRI scans, bypassing the need for segmentation masks or ground truth quantitative maps. Our approach utilizes a self-supervised learning framework that incorporates the MR-Zero simulation engine, ensuring the production of sharper, contrast-accurate images that strictly adhere to MRI physics. The Flow-based model yielded the best results, as its architectural design excels at preserving anatomical details of the brain. The developed framework enables physically grounded MRI data augmentation, significantly expanding training datasets while avoiding the financial costs and ethical constraints associated with volunteer scanning.

# Acknowledgments

This work was supported by the Ministry of Science and Higher Education of the Russian Federation (ProjectFSER-2025-0009).

# Data Availability Statement

The source code for training the flow-based model integrated with the differentiable MR-Zero MRI simulator is publicly available in the GitHub repository: https://github.com/KseniyaMD/Phantoms_generation_MR_Zero.

# References

1. Wegner, M., Gargioni, E., & Krause, D. (2023). Classification of phantoms for medical imaging. *Procedia CIRP*, *119*, 1140-1145.
2. B. Aubert-Broche, A. C. Evans, and D. L. Collins, “A new improved version of the realistic digital brain phantom, ” NeuroImage, vol. 32, no. 1, pp. 138–145, 2006. doi: 10.1016/j.neuroimage.2006.03.052.
3. Keenan, K. E., Jordanova, K. V., Ogier, S. E., Tamada, D., Bruhwiler, N., Starekova, J., ... & Hernando, D. (2024). Phantoms for Quantitative Body MRI: a review and discussion of the phantom value. *Magnetic Resonance Materials in Physics, Biology and Medicine*, *37*(4), 535-549.
4. S. Takahashi et al., “Comparison of vision transformers and convolutional neural networks in medical image analysis: a systematic review, ” J. Med. Syst., vol. 48, no. 1, p. 84, 2024. doi: 10.1007/s10916-024-02105-8.
5. Chen, L., Zhang, C., Yi, Y., Wang, Y., Song, Y., Yan, X., ... & Yang, G. (2026). A physics-driven neural network with parameter embedding for generating quantitative MR maps from weighted images. *Medical Physics*, *53*(3), e70394..
6. H. Lajous et al., “A fetal brain magnetic resonance acquisition numerical phantom (FaBiAN), ” Sci. Rep., vol. 12, p. 8682, 2022. doi: 10.1038/s41598-022-10335-4
7. Kumar, N. M., Fritz, B., Stern, S. E., Warntjes, J. M., Lisa Chuah, Y. M., & Fritz, J. (2018). Synthetic MRI of the knee: phantom validation and comparison with conventional MRI. *Radiology*, *289*(2), 465-477.

8. M. Goyal and Q. H. Mahmoud, “A systematic review of synthetic data generation techniques using generative AI, ” Electronics, vol. 13, no. 17, p. 3509, 2024.

9. Qiu, S., Wang, L., Sati, P., Christodoulou, A. G., Xie, Y., & Li, D. (2024). Physics-guided self-supervised learning for retrospective T1 and T2 mapping from conventional weighted brain MRI: Technical developments and initial validation in glioblastoma. *Magnetic Resonance in Medicine*, *92*(6), 2683-2695.

10. Wang, Ke, et al. "High-fidelity direct contrast synthesis from magnetic resonance fingerprinting." *Magnetic Resonance in Medicine* 90.5 (2023): 2116-2129.

11. Gulani, Vikas, et al. "Towards a single-sequence neurologic magnetic resonance imaging examination: multiple-contrast images from an IR TrueFISP experiment." *Investigative radiology* 39.12 (2004): 767-774.

12. Warntjes, J. B. M., et al. "Rapid magnetic resonance quantification on the brain: optimization for clinical usage." *Magnetic Resonance in Medicine: An Official Journal of the International Society for Magnetic Resonance in Medicine* 60.2 (2008): 320-329.

13. Cheng, Cheng-Chieh, et al. "Multipathway multi-echo (MPME) imaging: all main MR parameters mapped based on a single 3D scan." *Magnetic resonance in medicine* 81.3 (2019): 1699-1713.

14. Dayarathna, Sanuwani, et al. "Deep learning based synthesis of MRI, CT and PET: Review and analysis." *Medical image analysis* 92 (2024): 103046.

15. Hamilton, Jesse I., and Nicole Seiberlich. "Machine learning for rapid magnetic resonance fingerprinting tissue property quantification." *Proceedings of the IEEE* 108.1 (2019): 69-85.

16. Golbabaee, Mohammad, et al. "Geometry of deep learning for magnetic resonance fingerprinting." *ICASSP 2019-2019 IEEE International Conference on Acoustics, Speech and Signal Processing (ICASSP)*. IEEE, 2019.

17. S. Wang, H. Ma, J. A. Hernandez-Tamames, S. Klein, and D. H. J. Poot, “qMRI diffuser: quantitative T1 mapping of the brain using a denoising diffusion probabilistic model, ” in Proc. MICCAI Workshop Deep Generative Models, Cham, Switzerland, Oct. 2024, pp. 129–138.

18. Sun, H., Wang, L., Daskivich, T., Qiu, S., Han, F., D'Agnolo, A., ... & Xie, Y. (2023). Retrospective T2 quantification from conventional weighted MRI of the prostate based on deep learning. *Frontiers in Radiology*, *3*, 1223377.

19. Lüpke, S., Yeganeh, Y., Adeli, E., Navab, N., & Farshad, A. (2024, October). Physics-informed latent diffusion for multimodal brain mri synthesis. In *International Conference on Medical Image Computing and Computer-Assisted Intervention* (pp. 198-207). Cham: Springer Nature Switzerland.

20. Isola, P., Zhu, J. Y., Zhou, T., & Efros, A. A. (2017). Image-to-image translation with conditional adversarial networks. In *Proceedings of the IEEE conference on computer vision and pattern recognition* (pp. 1125-1134).

21. W. Fan, J. Chen, and Z. Liu, "Hierarchy flow for high-fidelity image-to-image translation," 2023, arXiv:2308.06909.

22. Ravi, K. S., Geethanath, S., & Vaughan, J. T. (2019). PyPulseq: a python package for MRI pulse sequence design. *Journal of Open Source Software*, *4*(42), 1725.

23. B. Aubert-Broche, M. Griffin, G. B. Pike, A. C. Evans, and D. L. Collins, "Twenty new digital brain phantoms for creation of validation image data bases," IEEE Trans. Med. Imag., vol. 25, no. 11, pp. 1410–1416, Nov. 2006, doi: 10.1109/TMI.2006.883453.

24. V. C. Obmann, M. Ardoino, J. Klaus, et al., "MRI extracellular volume fraction in liver fibrosis—a comparison of different time points and blood pool measurements," J. Magn. Reson. Imaging, 2024, doi: 10.1002/jmri.29000.

25. T. Diekhoff, D. Deppe, D. Poddubnyy, et al., "Characterization of bone marrow lesions in axial spondyloarthritis using quantitative T1 mapping MRI," Skeletal Radiol., vol. 53, no. 7, pp. 1295–1302, 2024, doi: 10.1007/s00256-023-04196-0.

26. Brui, Ekaterina A., et al. "Comparative analysis of SINC-shaped and SLR pulses performance for contiguous multi-slice fast spin-echo imaging using metamaterial-based MRI." *Magnetic Resonance Materials in Physics, Biology and Medicine* 34.6 (2021): 929-938.

27. C. Castillo-Passi et al., "KomaMRI. jl: An open-source framework for general MRI simulations with GPU acceleration," Magn. Reson. Med., vol. 90, no. 1, pp. 329–342, 2023.

28. S. J. Meara and G. J. Barker, "Evolution of the longitudinal magnetization for pulse sequences using a fast spin-echo readout: application to fluid-attenuated inversion-recovery and double inversion-recovery sequences," Magn. Reson. Med., vol. 54, no. 1, pp. 241–245, 2005.

29. A. Loktyushin et al., "MRzero-automated discovery of MRI sequences using supervised learning," Magn. Reson. Med., vol. 86, no. 2, pp. 709–724, 2021.
30. X. Huang and S. Belongie, “Arbitrary style transfer in real-time with adaptive instance normalization,” in Proceedings of the IEEE International Conference on Computer Vision, 2017, pp. 1501–1510
31. G. P. Renieblas, A. T. Nogués, A. M. González, N. Gómez-Leon, and E. G. Del Castillo, "Structural similarity index family for image quality assessment in radiological images," J. Med. Imaging, vol. 4, no. 3, p. 035501, 2017.
32. A. Hore and D. Ziou, "Image quality metrics: PSNR vs. SSIM," in Proc. 20th Int. Conf. Pattern Recognit., Aug. 2010, pp. 2366–2369.
33. Kaufmann, Timothy J., et al. "Consensus recommendations for a standardized brain tumor imaging protocol for clinical trials in brain metastases." *Neuro-oncology* 22.6 (2020): 757-772.
34. Moya-Sáez, E., Peña-Nogales, Ó., de Luis-García, R., & Alberola-López, C. (2021). A deep learning approach for synthetic MRI based on two routine sequences and training with synthetic data. *Computer Methods and Programs in Biomedicine*, *210*, 106371.
35. Van Lune, Jelmer, et al. "Quantitative Mapping from Conventional MRI Using Self-Supervised Physics-Guided Deep Learning: Applications to a Large-Scale, Clinically Heterogeneous Dataset." Medical Image Analysis, vol. 115, 2027, p. 104295.
36. Dang, Hoai Nam, et al. "MR-zero meets RARE MRI: joint optimization of refocusing flip angles and neural networks to minimize T2-induced blurring in spin echo sequences." *Magnetic Resonance in Medicine* 90.4 (2023): 1345-1362.
37. Weinmüller, Simon, et al. "MR-zero meets FLASH–controlling the transient signal decay in gradient-and RF-spoiled gradient echo sequences." *Magnetic Resonance in Medicine* 93.3 (2025): 942-960.
38. Liu, Xiaofeng, et al. "Dual-cycle constrained bijective vae-gan for tagged-to-cine magnetic resonance image synthesis." *2021 IEEE 18th International Symposium on Biomedical Imaging (ISBI)*. Ieee, 2021.

39. Finck, Tom, et al. "Uncertainty-aware and lesion-specific image synthesis in multiple sclerosis magnetic resonance imaging: a multicentric validation study." *Frontiers in neuroscience* 16 (2022): 889808.

40. Osuala, Richard, et al. "Data synthesis and adversarial networks: A review and meta-analysis in cancer imaging." *Medical Image Analysis* 84 (2023): 102704.

41. Koetzier, Lennart R., et al. "Generating synthetic data for medical imaging." *Radiology* 312.3 (2024): e232471.

Figure 1. Overview of the proposed training pipeline. The pipeline takes weighted images (T1-, T2-, PD-w) as input to predict parametric maps via a neural network (VAE, GAN, or Flow-based). An MR simulator (Analytical or Numerical) then reconstructs weighted images from these maps. The model is optimized by minimizing the reconstruction loss between the input and the generated weighted images.

Figure 2. Overview of the multimodal variational autoencoder framework. T1-, T2-, and PD-weighted MR images are independently encoded into unimodal latent distributions using a shared convolutional encoder conditioned on pulse sequence parameters. The distributions are fused using the product-of-experts approach, and a convolutional decoder generates parametric T1, T2, and PD maps. These maps are subsequently used with an analytical MR signal model to reconstruct the corresponding weighted MR images. Training combines reconstruction and perceptual losses, PatchGAN adversarial loss, and KL-divergence regularization.

Figure 3. Flow-based model training scheme. The model consists of two Hierarchical Coupling Layers. It takes a three-channel array of weighted images (T1-, T2-, PD-w) as input, along

with the mean and variance parameters derived from a single sample of synthetic parametric maps using Style Net. The model synthesizes parametric maps (T1, T2, PD), which are then fed into an MR signal model to generate new weighted images for Mean Square Error (MSE) loss calculation.

Figure 4. GAN training scheme. The generator takes weighted images as input and generates parametric maps, which are then fed into an MR signal model to produce new weighted images. These generated weighted images are passed to the discriminator. The discriminator is trained using Binary Cross-Entropy (BCE) loss, while the generator is trained using Mean Absolute Error (MAE) calculated between the original and generated weighted images.

Figure 5. Examples of parametric maps: synthetic ground truth maps and those generated using VAE, GAN, Flow-based models.

Figure 6. Model results: A – images reconstructed using an analytical MR signal and B – generated parametric maps. T1(s) - T1 relaxation time in seconds, T2(s) - T2 relaxation time in seconds, PD(a.u.) - proton density in arbitrary units, T1w - T1-weighted image, T2w - T2-weighted image, PDw - PD-weighted image.

Figure 7. (A) Examples of parametric maps generated by the flow-based model under three training settings. HFModel1 – the model pre-trained on synthetic data using an analytic MR signal model, HFModel2 – the model pre-trained on synthetic and finetuned on real data using an analytic MR signal model, HFModel3 – the model pre-trained on synthetic and finetuned on real data using MR-Zero simulator. (B) Comparison between original TSE images and images generated using the flow-based model and the MR-Zero simulator. The flow-based model was trained using the MR-Zero simulator. HFModel1 – the model pre-trained on synthetic data using an analytic MR signal model, HFModel2 – the model pre-trained on synthetic and finetuned on real data using an analytic MR signal model, HFModel3 – the model pre-trained on synthetic and finetuned on real data using MR-Zero simulator. (C) Simulation of different brain MRI contrasts from a single digital phantom. From left to right: original T1-weighted TSE image and MR-Zero simulations of T1-weighted 2D FLASH (TR/TE/FA = 250ms/5ms/70°), turbo-FLASH-2D (TR/TE/TI/FA = 2100ms/4ms/1100ms/12°), and FLAIR-2D (TR/TE/TI/FA/ETL = 9000ms/110ms/2200ms/90°/180°/14). All simulated images were generated using a 256 × 256 matrix.

Table 1. The mean absolute error (MAE) between the original synthetic parametric maps and those generated using different models.

Table 2. Performance of VAE, GAN, and Flow-based models on the real-world test set, trained using an analytical MR signal model. MS-SSIM and PSNR metrics were calculated between input images and those reconstructed via the analytical MRI signal model using the generated maps. The results are presented before fine-tuning on a real dataset and after it.

Table 3. Results for the flow-based model fine-tuning. Pre-training was conducted on synthetic data using an analytical MR signal model, followed by fine-tuning on real-world data using either analytical model or the MR-Zero simulator. Testing was performed on a real-world dataset.